# WHY PATTERNS APPEAR SPONTANEOUSLY IN DISSIPATIVE SYSTEMS?

**K. Staliûnas**

Physikalisch-Technische Bundesanstalt, D-38023 Braunschweig, Germany

Abstract

It is proposed that the spatial (and temporal) patterns spontaneously appearing in dissipative systems maximize the energy flow through the pattern forming interface. In other words – the patterns maximize the entropy growth rate in an extended conservative system (consisting of the pattern forming interface and the energy bathes). The proposal is supported by examples of the pattern formation in different systems. No example contradicting the proposal is known.

## INTRODUCTION

Many different pattern-forming systems are known to exist in the nature, also in controllable laboratory experiments. A common feature for pattern forming systems is their openness, or dissipativity; this is why the spontaneously appearing patterns are also called by "dissipative patterns" [1]. Openness also mean, that a system is out of the thermal equilibrium, in contrary to conservative systems, where no patterns spontaneously appear, but instead the energy equipartition occurs, leading to most symmetrical, maximum entropy states.

One therefore defines the pattern as a state of matter with a spontaneously broken space symmetry [1]. (The spontaneous onset of periodic or chaotic oscillations in spatially confined system can be also considered as formation of "temporal" pattern, where not the space symmetry but the symmetry of time is broken.) The spatial or temporal pattern formation is the process, when the symmetry breaks, and more asymmetric, more unprobable, and more "especial" formations occurs.

Formation of rolls and hexagons in Rayleigh-Benard convection [2] is a seminal example of patterns in controllable laboratory experiment. Self-sustained oscillations in chemical reactions illustrate the symmetry breaking of space and time [3]. These elementary laboratory patterns and elementary evolution processes (the transients) can be considered as building bricks of more complicated patterns, and more complicated evolution processes observed in the nature. One can consider the complicated patterns in the nature as elementary patterns multiply nested within the other elementary patterns.

Some universal pattern formation principles are now understood and mathematically described. It is known, that the necessary condition for the occurrence of Turing patterns is the existence of unstable modes with nonzero characteristic spatial (temporal) scale of modulation [4]. The responsible physical mechanisms can be different: well known is the mechanism of "local activation and lateral inhibition" being at the root for patterns in many biological and chemical systems [5]. The other universal mechanism is also known to be responsible for the appearance of patterns in parametrically excited (Faraday type [6]) systems: when the system is (parametrically) excited at a frequency different than its internal resonance frequency [7].

Turing mechanism can be considered as a primary stage of pattern formation. The next stage would mean the secondary instabilities of ideal Turing patterns. Some universal principles are also known governing the pattern formation in this second stage: the ideal Turing patterns can show Eckhause, Zig-zag, or other instabilities, or defects, universally described e.g. by means of amplitude and phase equations [8].

Despite the progress of the last decades, one general question remains open: does there exists an universal functional whose maximization can be associated with the occurrence of patterns [1]? The functional, whose maximization could be associated with the formation of primary Turing patterns, as well as with the secondary instabilities responsible for various zigzags and defects of ideal Turing patterns. Or, generally speaking, the functional, whose maximization could be associated with the evolution towards more complicated, more unprobable, more symmetry broken, and lower entropy states in various pattern-forming systems?

No such universal evolution principle for the dissipative systems has been formulated up to now. From the other side, the "evolution" of the conservative systems has been understood and mathematically described already long time ago. The evolution is governed there by the second foundation of thermodynamics, stating that every closed systems evolve in such a way, that its entropy constantly grows, and reaches the maximum value at a thermal equilibrium. As a consequence the final macroscopic state of a closed system is maximally probable, and maximally symmetric one, because it contains the largest number of microscopic states.

Can there be formulated an universal evolution principle for dissipative systems, perhaps stating that the entropy decreases, and the evolution occurs, leading to some pattern as a "most unprobable" state?

The idea is proposed in the letter, that the pattern formation and the hypothetical evolution principle is not a fundamental principle by itself, but serves to the more general principle – the second foundation of thermodynamics. It is proposed that all spatial patterns (and all dynamical regimes) in nonequilibrium system occur with a single purpose: in order to optimize the entropy growth of an extended conservative system, and to accelerate maximally the burn-out of entropy resources. By an extended closed system here is assumed: the nonequilibrium system, where pattern formation occurs (the pattern forming interface), plus the energy sources and sinks supplying the energy to the interface and driving it from thermal equilibrium.

The statement of the letter can be illustrated on many examples of pattern forming systems. Some examples:

i.) Rayleigh-Benard convection [2]. When the temperature difference is relatively small, then the heat is transferred from the lower (hot) to the upper (cold) plate by the mechanism of thermal conduction, therefore the heat transfer is linearly proportional to the temperature difference. However when the temperature difference exceeds a particular threshold the convection rolls appear, leading to a sharp increase of the heat transfer rate at the very onset of convection.

The onset of convection rolls can be considered as the optimization of the heat transfer rate, also of the rate of growth of the entropy in the extended system (including fluid interface and the both thermal reservoirs). The system prefers a pattern which ensures the quickest burn-out of the entropy resources, thus enables the fastest route to the thermal equilibrium. The system switches spontaneously from the unstable homogeneous conduction state, to the stable unhomogeneous (pattern) state in order to maximize the entropy production in total ii.) Stirred chemical reactions [3]. The onset of self-sustained oscillations also helps to optimize the averaged rate of the chemical reaction. When the stationary chemical reaction becomes unstable and periodic pulsation occurs, the average fusion rate sharply increases. The system switches spontaneously from the unstable stationary state, to the stable periodic state in order to optimize of the chemical fusion, thus to ensure the fastest way of reaching the thermal equilibrium among the all chemical components participating in reaction.

iii.) Parametrically driven dispersive systems [7]. The patterns appear, when the spatially extended system is excited by the periodical modulation of a parameter of the system with a frequency slightly larger than twice of its resonance frequency. The spatial wavenumber of the pattern is such, that the frequency mismatch is compensated due to dispersive properties of the material.

When no pattern is excited, the external energy is only weakly dissipated for off-resonance system. To increase the energy transfer the spatial pattern spontaneously appears. The spatial pattern plays a role of the "bridge" between two frequencies – these of excitation and dissipation. Due to the pattern the system consumes effectively the energy of excitation and converts it into the thermal energy. The pattern here again enables a quickest growth of entropy of a system including energy source, thermal bath, and the pattern forming interface.

iv) The biological evolution. The driving mechanism of biological evolution is a natural survival of the most adapted species. It means nothing, but the survival of such species which can most effectively consume the resources of the environment. Thus the Darwin evolution maximizes a total consumption of resources, thus maximizes an entropy growth of the corresponding extended system.

This list of examples can be continued without end. Patterns of turbulence sets in because they enable the most effective energy transfer through the spatial scales. And finally the industry and civilization develops perhaps because they maximize the burn-out of the entropy resources of the earth. Absolutely in all cases the patterns occur in order to burn-out as fast as possible the sources creating them. No example of pattern is known, which would serve to the opposite purpose: which would isolate the bathes of energy one from another, thus slow down the entropy growth. Every spontaneously appearing pattern increases the coupling between the bathes, thus maximizes the energy exchange and the growth of entropy.

Fig.1 illustrates schematically the chain of hierarchical pattern forming systems. The Rayleigh-Benard convection cell, reaction diffusion system, or parametrically excited system can be sketched schematically by only one "pattern formation" interface between two thermal bathes, thus is more straightforwardly tractable by means of nonequilibrium thermodynamics. More complicated pattern forming system, like biological ones, consists of more pattern forming interfaces, thus the patterns from n-th hierarchical level not necessary feed directly from the thermal bathes, but supposingly via contact with different nonequilibrium patterns in (n-1)-th hierarchical level.

The first order patterns seem to be the Turing ones according to the suggested scheme in Fig.1. The second order patterns should appear when two Turing patterns characterized by different spatial wavenumbers are involved. One such example with two Turing patterns involved is a zigzag instability. Therefore the zig-zags (or in general the secondary instabilities, or defects of ideal Turing patterns) can be considered as an example of second order pattern.

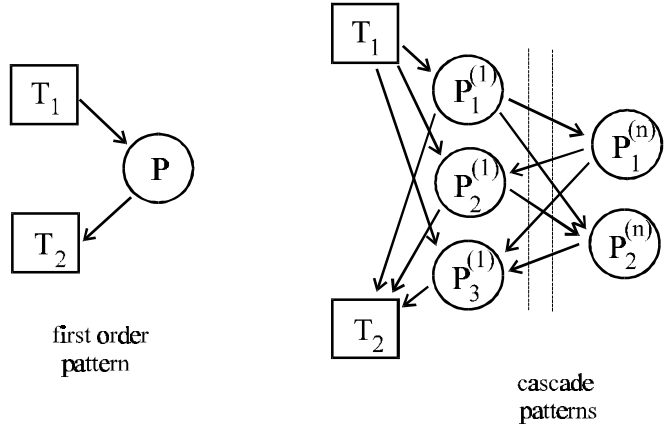

Fig.1: Illustration of elementary pattern (at the left) and of higher order patterns (at the right). $T_i$ – denote the energy bathes, $P_i$ – the pattern forming interfaces.

As a conclusion: the pattern formation principle in dissipative systems is proposed in the letter, basing on the maximization of entropy growth of extended conservative system. Every pattern in dissipative system appears with the purpose to maximize the energy flow through the system, or in other words - to maximize the entropy growth rate of the extended system. The proposed principle is supported by numerous examples of pattern forming systems.

No example of a system contradicting the suggested principle is known.

The proposed principle of the maximum entropy growth rate seemingly can not be proven in general case. The proof can not be performed in the frame of general mathematical models describing pattern formation (in form of PDEs such as Swift-Hohenberg equation, Ginzburg-Landau equations and their modifications), because the physical information on the energy sources are lost in deriving these models. Only some information on pattern forming interface remains in these general mathematical models. But as the formulation of the pattern formation principle extends the frames of pattern forming interface, the above mentioned general mathematical models are insufficient.

**Kodėl dariniai atsiranda spontaniškai disipatyviose sistemose?**

K. Staliûnas

Straipsnyje keliama hipotezė, kad erdviniai (ir laikiniai) dariniai, spontaniškai atsirandantys disipatyviose sistemose maksimizuoja energijos srautą[1] per darinius formuojantá interfeisą. Kitaip tariant – dariniai maksimizuoja entropijos augimo greitá išplėstinėje konservatyvioje sistemoje (sudarytoje is darinius formuojančio interfeiso ir energijos virsmø). Đià hipotezæ patvirtina eilë pavyzdþ iø apie dariniø formavimasá ávairiose sistemose. Be to, joks þ inomas pavyzdys nepaneigia ðios hipotezës.